\documentclass[aps,prl,twocolumn,reprint,groupedaddress,superscriptaddress]{revtex4-1}

\usepackage{graphicx}
\usepackage{amsmath}

\begin{document}

%\preprint{}

\title{Realization of strong coupling between deterministic single-atom arrays \\
and a high-finesse miniature optical cavity}


\affiliation{State Key Laboratory of Quantum Optics and Quantum Optics Devices,and
Institute of Opto-Electronics, Shanxi University, Taiyuan 030006,
China}
\affiliation{Collaborative Innovation Center of Extreme Optics, Shanxi University,
Taiyuan 030006, China}

\author{Yanxin~Liu}
\thanks{These authors contributed equally to this work.}
\affiliation{State Key Laboratory of Quantum Optics and Quantum Optics Devices,and
Institute of Opto-Electronics, Shanxi University, Taiyuan 030006,
China}
\affiliation{Collaborative Innovation Center of Extreme Optics, Shanxi University,
Taiyuan 030006, China}

\author{Zhihui~Wang}
\thanks{These authors contributed equally to this work.}
\affiliation{State Key Laboratory of Quantum Optics and Quantum Optics Devices,and
Institute of Opto-Electronics, Shanxi University, Taiyuan 030006,
China}
\affiliation{Collaborative Innovation Center of Extreme Optics, Shanxi University,
Taiyuan 030006, China}

\author{Pengfei~Yang}
\affiliation{State Key Laboratory of Quantum Optics and Quantum Optics Devices,and
Institute of Opto-Electronics, Shanxi University, Taiyuan 030006,
China}
\affiliation{Collaborative Innovation Center of Extreme Optics, Shanxi University,
Taiyuan 030006, China}

\author{Qinxia~Wang}
\affiliation{State Key Laboratory of Quantum Optics and Quantum Optics Devices,and
Institute of Opto-Electronics, Shanxi University, Taiyuan 030006,
China}
\affiliation{Collaborative Innovation Center of Extreme Optics, Shanxi University,
Taiyuan 030006, China}

\author{Qing~Fan}
\affiliation{State Key Laboratory of Quantum Optics and Quantum Optics Devices,and
Institute of Opto-Electronics, Shanxi University, Taiyuan 030006,
China}
\affiliation{Collaborative Innovation Center of Extreme Optics, Shanxi University,
Taiyuan 030006, China}

\author{Shijun~Guan}
\affiliation{State Key Laboratory of Quantum Optics and Quantum Optics Devices,and
Institute of Opto-Electronics, Shanxi University, Taiyuan 030006,
China}
\affiliation{Collaborative Innovation Center of Extreme Optics, Shanxi University,
Taiyuan 030006, China}

\author{Gang Li}
\email{gangli@sxu.edu.cn}
\affiliation{State Key Laboratory of Quantum Optics and Quantum Optics Devices,and
Institute of Opto-Electronics, Shanxi University, Taiyuan 030006,
China}
\affiliation{Collaborative Innovation Center of Extreme Optics, Shanxi University,
Taiyuan 030006, China}

\author{Pengfei Zhang}
\email{zhangpengfei@sxu.edu.cn}
\affiliation{State Key Laboratory of Quantum Optics and Quantum Optics Devices,and
Institute of Opto-Electronics, Shanxi University, Taiyuan 030006,
China}
\affiliation{Collaborative Innovation Center of Extreme Optics, Shanxi University,
Taiyuan 030006, China}

\author{Tiancai Zhang}
\email{tczhang@sxu.edu.cn}
\affiliation{State Key Laboratory of Quantum Optics and Quantum Optics Devices,and
Institute of Opto-Electronics, Shanxi University, Taiyuan 030006,
China}
\affiliation{Collaborative Innovation Center of Extreme Optics, Shanxi University,
Taiyuan 030006, China}


\begin{abstract}
We experimentally demonstrate strong coupling between a one-dimensional (1D) single-atom array and a high-finesse miniature cavity. The atom array is obtained by loading single atoms into a 1D optical tweezer array with dimensions of 1$\times$11. Therefore, a deterministic number of atoms is obtained, and the atom number is determined by imaging the atom array on a CCD camera in real time. By precisely controlling the position and spacing of the atom array in the high finesse Fabry--Perot cavity, all the atoms in the array are strongly coupled to the cavity simultaneously. The vacuum Rabi splitting spectra are discriminated for deterministic atom numbers from 1 to 8, and the $\sqrt{N}$ dependence of the collective enhancement of the coupling strength on atom number $N$ is validated at the single-atom level. 
\end{abstract}


%\keywords{}

\maketitle

A strongly coupled cavity quantum electrodynamics (QED) system is a basic physical system for studying light-matter interactions \cite{CQED2005}, which not only is a test bed for studying fundamental physics but also provides powerful quantum resources for quantum information \cite{RevModPhys.73.565,RevModPhys.85.1083,RevModPhys.87.1379,RevModPhys.93.025005,RevModPhys.85.553}. Matured through single-atom control in the small cavity mode, vacuum Rabi splitting of a single atom has been observed \cite{PhysRevLett.68.1132, PhysRevLett.93.233603, PhysRevLett.94.033002}, which has provided great significance in quantum physics. As promising platforms to realize quantum networks \cite{Nature.Kimble2008Riew}, optical cavity QED systems have attracted intense interest. Research has been mainly focused on the interaction between single atoms and single photons. Many new quantum technologies and devices, e.g., single-photon sources and blockades \cite{Nature.rempe2003,science.1095232,PhysRevLett.89.067901,PhysRevLett.123.133602,Nature.rempe.2005,Nature.photonics.Dayan.2016,PhysRevLett.118.133604}, quantum interfaces \cite{science.1143835,PRXQuantum.2.020331}, quantum logic gates \cite{Nature.Rempe.2014,Nature.Rempe.2016,NaturePhysics.Dayan.2018,Science.Rempe.2020}, quantum measurements \cite{Nature.Reichel.2011,science.1246164,PhysRevLett.92.127902,Nature.Rempe.2021,NaturePhotonics.Emanuele.2021,PhysRevLett.126.253603}, and quantum routers \cite{Science.Dayan.2014,Science.Arno.2016,Science.Kimble.2008,PhysRevLett.101.203602,PhysRevLett.102.083601}, have been developed and investigated. Significantly, the demonstration of an elementary quantum network \cite{Nature.Rempe.2012} between two nodes with an individual atom in each cavity has brought a great leap forward for quantum networks. 

The multiatom cavity QED system, in which individual atoms can be discriminated and controlled, would be more interesting for both fundamental physics and applications. First, the cavity photon-mediated interactions between different atoms enrich the dynamics and complexity of the coupling diagrams for many-body physics research \cite{PhysRevLett.122.010405,Nature.Thompson.2020,Nature.Smith.2021}. Moreover, in the context of the recent progress in programmable arrays of atoms in quantum simulations and quantum computations \cite{RevModPhys.82.2313,Nature.Lukin.2021,Nature.Browaeys.2021,NaturePhysics.Endres.2020,Nature.Saffman.2022,Nature.Lukin.2022}, the development of a multiqubit module with optical links, which can process quantum information locally and interface qubits between atoms and photons, brings new perspectives for quantum networks and distributed quantum computation \cite{science.abi9917}.

However, building such a multiatom cavity QED system is quite challenging because of the stringent requirement on the position control of every atom to obtain steady and uniform strong coupling to a tiny cavity mode for each atom. To date, only two neutral atoms have been successfully controlled in the same mode of a Fabry--Perot (F-P) or nanophotonic cavity \cite{science.abi9917,PhysRevLett.111.100505,PhysRevLett.118.210503,PhysRevLett.114.023601}. A cavity QED system with one-dimensional (1D) atom arrays transversely integrated with a high finesse F-P cavity has also been recently presented \cite{PhysRevLett.128.083201, arXiv.2022, PhysRevA.105.043321}. However, the atoms are not uniformly coupled to the cavity mode. A system of 5 ions coupled to an F-P cavity has also been demonstrated \cite{PhysRevLett.116.223001}, but not in the strong coupling regime.  In this letter, we report strong coupling between 1D atom arrays and a miniature F-P cavity. The atom arrays are engineered to couple to the cavity simultaneously with a uniform coupling strength. Strong coupling of up to 8 atoms of an 11-tweezer array is demonstrated. Vacuum Rabi splitting can be discriminated from 1 to 8 atoms individually, and the $\sqrt{N}$-scaling of the collective enhancement of the coupling strength with atom number $N$ is validated for a deterministic number of atoms.

\begin{figure}%[htbp]
\centering
	\includegraphics[width=\columnwidth]{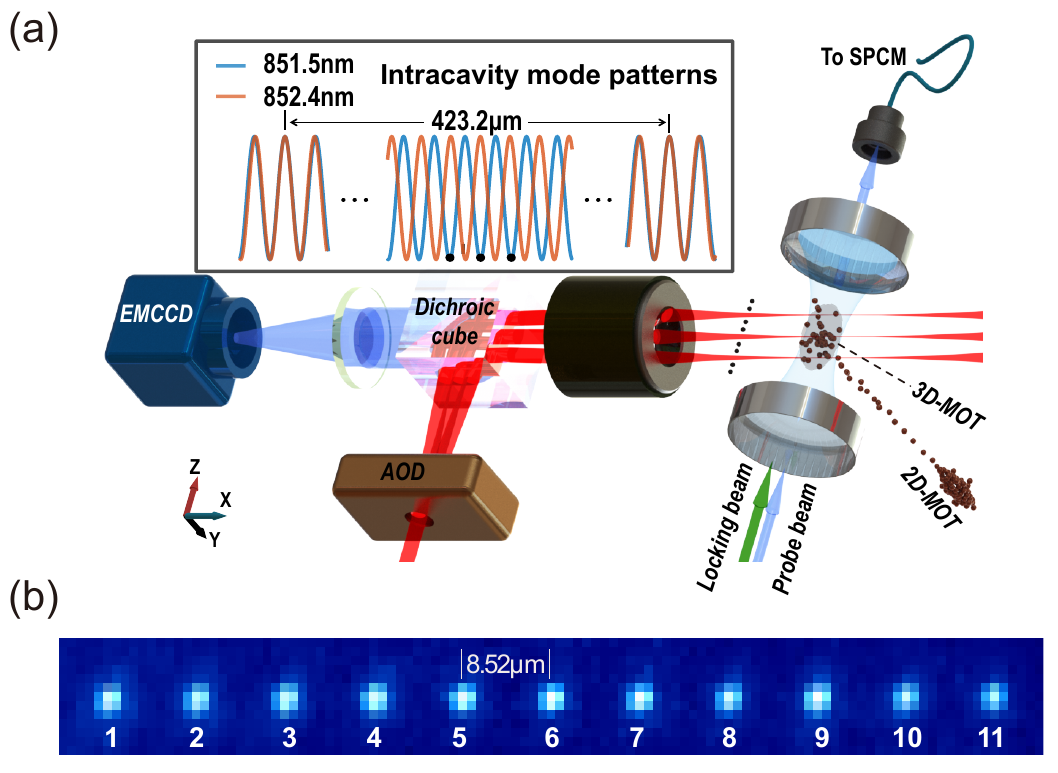}
\caption{\label{fig1} Scheme of the experiment. (a) Sketch of the essential part of the experimental setup. The F-P cavity is stabilized by an 851.5-nm locking beam, and another weak 852-nm probe beam (the corresponding intracavity mean photon number $\langle n \rangle =0.1$) is used to measure the spectrum. Inset: An illustration of the mode patterns of the lattice (blue) and 852-nm field (red). (b) Image of single atom arrays obtained by superimposing approximately 500 loading trials. The exposure time was set as 40 ms for every trial.}
\end{figure}

The experimental setup is illustrated in Fig. \ref{fig1}(a). The core of the setup is a miniature optical F-P cavity [picture can be found in the Supplemental Materials (SM)] \cite{SM}, which is composed of two high-reflectively coated mirrors with a curvature radius of 100 mm. The mirrors have transmittances of 4.9 and 84.9 ppm at 852 nm. The length of the cavity is fixed at 1.27 mm to accommodate the atom array while maintaining strong coupling for individual atoms. The waist of the TEM${_{00}}$ mode and finesse of the cavity are 46 $\mu$m and  $5.7 \times {10^4}$, respectively. Thus, the cavity QED parameters for individual cesium atoms in our system are $({g_0},\kappa ,\gamma ) = 2\pi  \times (3.2,1.0,2.6)$  MHz, where ${g_0}$ denotes the theoretical maximum coupling strength between a cesium atom (for transition $ |g\rangle \equiv 6{S_{1/2}}|F = 4,{m_F} = 4\rangle  \leftrightarrow |e\rangle \equiv 6{P_{3/2}}|F = 5,{m_F} = 5\rangle$) and the cavity TEM${_{00}}$ mode. $\kappa$ and $\gamma$ are the decay rates for the cavity and atom, respectively. The cooperative coefficient is $C={g_0}^2/2\kappa\gamma=1.9$, which means that our system is a strongly coupled cavity QED system for a single atom when the position can be controlled precisely at the antinode of the cavity standing-wave mode. 

The details of the experimental apparatus, including the optical cavity, the vacuum system, and the cavity locking scheme, can be found in SM \cite{SM}. The whole cavity system is placed inside a high-vacuum glass cell with inner dimensions of 20 mm $\times$ 25 mm on the cross section. The length of the cavity is actively stabilized to the cesium transition line $ |g\rangle  \leftrightarrow |e\rangle $ (the resonant wavelength is approximately 852.356 nm) by an auxiliary locking laser at 851.5 nm (3 free spectral ranges off the atomic transition), whose frequency has been locked to the cesium transition line via a transfer cavity. The locking laser also forms a lattice with positive potentials along the cavity axis. Due to the relatively long cavity length, a small magneto-optical trap (MOT) can be built directly inside the F-P cavity to accumulate the atoms emitted from the first-stage two-dimensional MOT. The atomic ensemble has a diameter of 150 $\mu$m and an atom number of approximately $10^5$. The temperature is approximately 15 $\mu$K after polarization gradient cooling.

The optical tweezer array is generated by strongly focusing a 1D laser beam array with dimensions of $1\times11$ by a homemade high-numerical-aperture objective with $NA=0.4$ and focal length $f=28.8$ mm \cite{RSI.Li.2020}. The laser beam array comes from the diffraction of an acousto-optic deflector (AOD, DTSX, AA Opto Electronic) driven by a multitone radio frequency (RF) signal. Every tweezer has a waist radius of 1.81 $\mu$m, which ensures that only a single atom is loaded by the light-assisted collision process  \cite{PhysRevLett.89.023005}. The tweezer array is projected into the cavity transversely from the outside of the vacuum glass cell and the orientation is along the cavity axis (Z-axis). The optical tweezers load single atoms directly from the precooled cesium atom ensemble. The fluorescence of the loaded single atoms is collected by the same objective, separated from the trapping beam by a dichroic beam splitter cube, and eventually imaged on an electron-multiplying CCD (EMCCD) camera. Figure \ref{fig1}(b) shows an average picture of the single atoms trapped by the tweezer array.

\begin{figure}%[htbp]
\begin{center}
\includegraphics[width= \columnwidth]{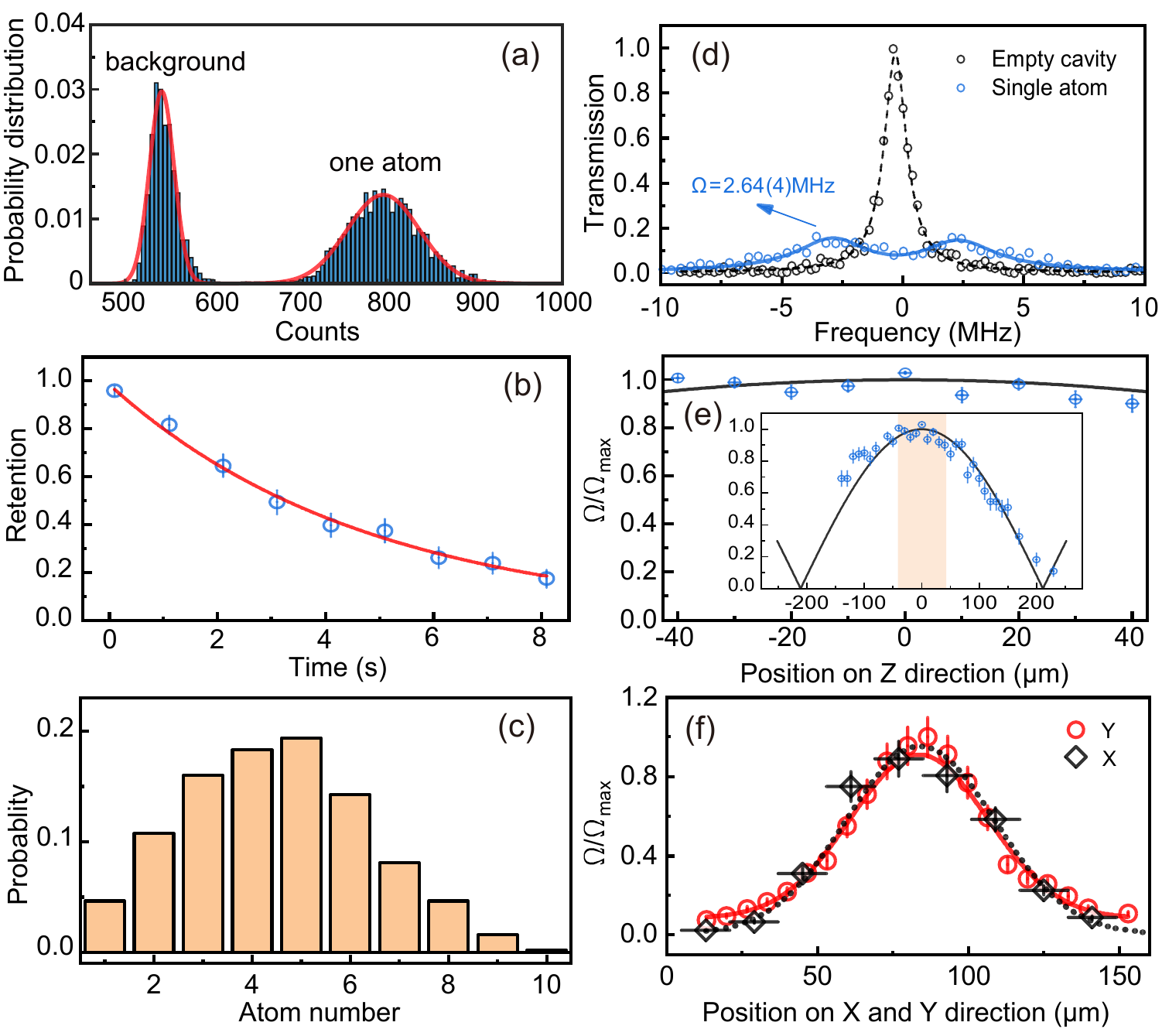}
\end{center}

\caption{\label{fig2} (a) Typical histogram of the electron counts of the fluorescence from the No. 6 optical tweezer in Fig.\ref{fig1}(b) on the EMCCD camera for 5000 trials of atom loading. The red line is the fitting with a bimodal Gaussian function. (b) Atom retention versus  atom holding time. The exponential fitting gives a characteristic atom lifetime of 4.8(1) s. (c) Probability of the number of single atoms loaded into all 11 optical tweezers. The probabilities are counted for 500 trials of atom loading, giving an average atom number of approximately 5. (d) Typical spectra during optimization. The light blue circles represent the experimental data with one tweezer on. The coupling strength $\Omega $ is determined by fitting with Formula (\ref{eq1}). The spectrum of the empty cavity is also shown for comparison (black circles). (e) Optimization of the position of the atom on the Z axis. The main figure shows the variance of the coupling strength in the position range from $-42.6$ to $42.6$ $\mu$m (shaded area of the inset figure), in which the optical tweezer arrays are placed. The inset shows the variance of coupling strength in a larger range. The black solid line is the theoretical coupling strength between the cavity and atom when the atom is trapped in different nodes of the intracavity lattice. The light blue circles are the measured atom-cavity coupling versus the atom position with only one tweezer on. (f) Dependence of the atom-cavity coupling on the X and Y axis positions. All eleven tweezers are used for these measurements. The fittings by Gaussian functions give mode waists of 48.0(2.8) and 45.7(1.7) $\mu$m along the X and Y axes, respectively.}
\end{figure}

Figure \ref{fig2}(a) shows a typical histogram of the fluorescence from one of the tweezers recorded by the EMCCD camera. The bimodal structure of the count distribution indicates that each time, only one atom is loaded into one tweezer, and the loading probability is approximately 57\%. The average lifetime for the trapped individual atom is measured to be approximately 4.8(1) s [as shown in Fig. \ref{fig2}(b)] when the tweezer overlaps with the intracavity lattice. The lifetime is limited by the heating due to the variance of the position overlap between the optical tweezer and the intracavity lattice. Figure \ref{fig2}(c) shows the atom number distribution in all 11 optical tweezers. From the single-atom loading probability (57\%), we expect the average atom number for all 11 tweezers to be approximately 6.3. However, we only obtain a value of 5 from Fig. \ref{fig2}(c). The reason is that several tweezers on the edge do not exactly overlap with the atomic ensemble in the measurement. If all the tweezers overlap well with the atomic ensemble, then the average atom number could reach the expected result.

The challenge of the experiment is to control the position of every atom to reach maximum and steady coupling to the cavity. To obtain this condition, the position of each tweezer should not only exactly overlap  with an antinode of the cavity standing-wave mode in the Z direction but also be in the center of the mode profile in both the X and Y directions. However, since the size of the optical tweezer is much larger than the structure of the standing wave, the atom cannot be confined around the small antinode region by the optical tweezer alone. With the aid of the blue-detuned lattices induced by the 851.5-nm locking laser the problem can be resolved. 

When the blue lattice is taken into account, the atom trapped in the tweezer will be pushed to a node of the lattice. The node of the lattice overlaps perfectly with the antinode of the cavity mode at the center of the 852-nm mode where the coupling between the atom and the cavity is maximum. The node of the lattice will gradually displace from the antinode of the 852-nm mode and totally mismatch with each other at the offset position with a distance of 216.6 $\mu$m to the cavity center, where the atom decouples to the cavity. The overlap repeats for every 423.2 $\mu$m along the Z direction. The solid line in Fig. \ref{fig2}(e) presents the theoretical prediction of the coupling strength of the atom when it is trapped in different sites of the lattice along the Z direction.

To verify the coupling pattern of the atom, only one tweezer is used to load the atom and test the coupling. The tweezer is switched on by driving the AOD with a single tone RF signal of 79.8 MHz. The position of the tweezer trap in the Z direction can be scanned by either a motorized stage or the RF driving frequency applied to the AOD. Here, it is scanned step by step by the motorized stage. At every scanned spatial point, a single atom is loaded into the tweezer and the atom-cavity coupling strength is checked by measuring the vacuum Rabi splitting spectrum. A typical spectrum is shown as the light blue data points in Fig. \ref{fig2}(d). The coupling strength $\Omega$ can be obtained by fitting the data with the theoretical transmission spectrum \cite{SM}
\begin{equation}
T = \frac{{{\kappa ^2}({\gamma ^2} + \Delta _{pa}^2)}}{{{{({\Omega ^2} - \Delta _{pa}^2 + {\Delta _{ca}}{\Delta _{pa}} + \gamma \kappa )}^2} + {{(\kappa {\Delta _{pa}} + \gamma {\Delta _{pa}} - \gamma {\Delta _{ca}})}^2}}}, \label{eq1}
\end{equation}
where ${\Delta _{ca}}$ (${\Delta _{pa}}$) is the frequency detuning between the cavity (probe) and atom. $\Omega $ is the coupling strength, which equals $g$ for a single atom. ${\Delta _{ca}}$ can also be determined by fitting. The relative coupling strength ${\Omega  \mathord{\left/
 {\vphantom {\Omega  {{\Omega _\text{max }}}}} \right.
 \kern-\nulldelimiterspace} {{\Omega _\text{max }}}} $, where ${\Omega _\text{max}}$ is the maximum in one scan trial, versus the position on the Z axis is shown as the blue circles in Fig. \ref{fig2}(e), which agrees well with the theoretical prediction from the overlap between the blue lattice and 852-nm cavity mode.

Thus, maximum and steady coupling can be naturally achieved as long as the 11 tweezers are placed around the center of the cavity where the node of the lattice and the antinode of the 852-nm mode coincide well in space [as shown in the inset of Fig. \ref{fig1}(a)]. The 11 tweezers are switched on with the center RF frequency fixed at 79.8 MHz. The distance between the neighboring tweezers is set as 8.52 $\mu$m by setting the spacing of the multitone RF frequency as 1.16 MHz. The spatial offset of the tweezers at the edge is $\pm 42.6$ $\mu$m to the cavity center. The coupling of the atom in the edge tweezer is 95\% of the one in the center in theory. The theoretical variance of the coupling for all 11 atoms in the tweezer array is only $\pm 2.5\%$. From the data measured in Fig. \ref{fig2}(e) we obtain that the variance of the measured coupling strength is within $\pm 4\%$.
 
\begin{figure}%[htbp]
\begin{center}
\includegraphics[width=\columnwidth]{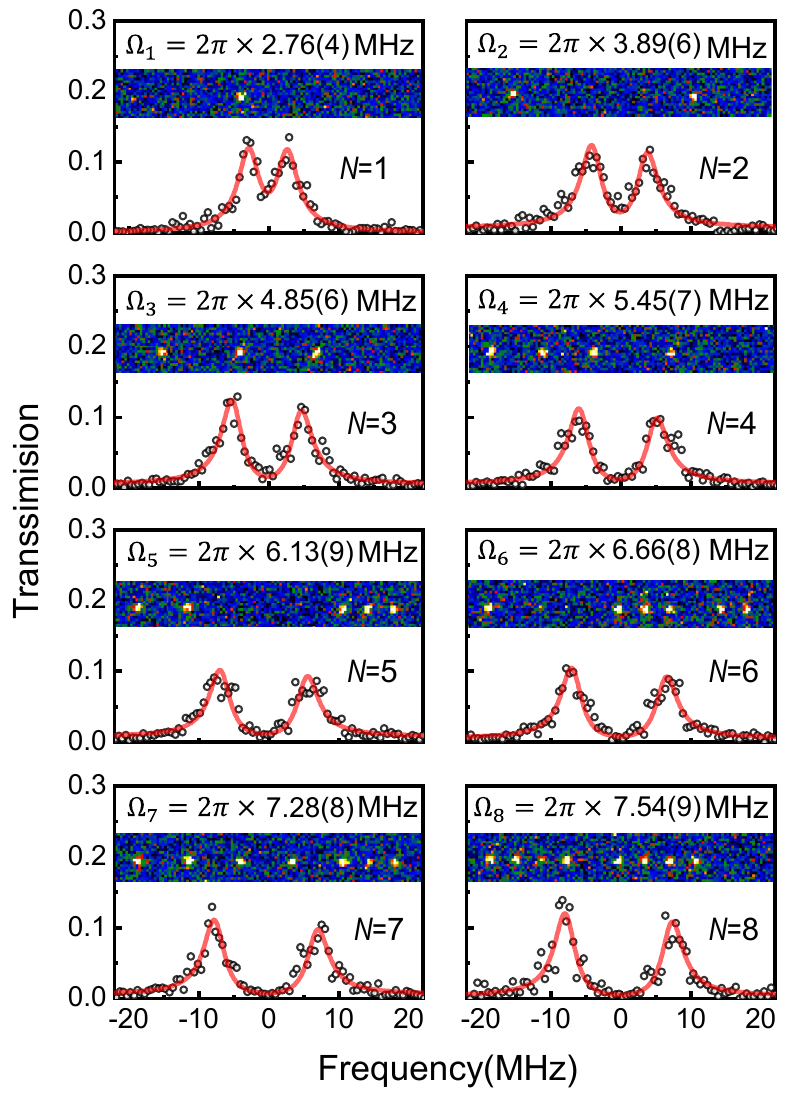}
\end{center}
\caption{\label{fig3} Vacuum Rabi splitting with a deterministic atom number from $N=1$ to $N=8$. The single shot images of the trapped atoms are shown as the inset picture, which are used to precisely count the atom number. The experimental data (black circles) are fitted (red lines) by Formula (\ref{eq1}) to determine ${{\Omega _N}}$.}
\end{figure}

After the Z-positions of the atomic array have been optimized at the maximum coupling spots, the positions in the X and Y directions are also scanned with all eleven tweezers on. The measured coupling coefficients versus the X and Y positions are displayed in Fig. \ref{fig2}(f), where the results follow a Gaussian mode profile well. The fitting of the results by Gaussian functions gives mode waists in the X and Y directions of 48.0(2.8) and 45.7(1.7) $\mu$m, respectively, which are in good agreement with those calculated from the geometry of the F-P cavity. Therefore, by setting the positions in the X and Y directions to the maximum coupling spots, we can eventually optimize and realize strong coupling between deterministic atom arrays and the F-P cavity.

In principle, our cavity QED system can realize strong coupling between the F-P cavity and atoms with a deterministic number less than 11. Here, we demonstrate coupling between the cavity and atom arrays with atom numbers from 1 to 8. The measured vacuum Rabi spectra in the transmission and images of the atom arrays are depicted in Fig. \ref{fig3}. The atom number is exactly determined from the EMCCD image and the coupling strength $\Omega$ is extracted by fitting with Formula (\ref{eq1}). The measured vacuum Rabi splitting of a single atom is $2g={\rm{2}}\pi  \times 5.52(8)$ MHz, which is approximately 87\% of the maximum theoretical value of ${\rm{2}}\pi  \times 6.32 $ MHz. The difference mainly comes from the imperfect state initialization efficiency ($\sim$91\%). The average due to residual motion of the atom will also produce a smaller $g$. The unequal height of the two normal splitting peaks in Fig. \ref{fig3} mainly comes from the uneven ${\Delta _{ca}}$ in different tweezers. Since we leave all the tweezers on with a shallow trap depth ($\sim$0.1 mK) during the measurement, the light shifts of atoms in different tweezers are uneven due to the small variance in the trap shapes and intensities. The ${\Delta _{ca}}$ values extracted from the data fitting are within the range of 0--$-0.4$ MHz for all the subfigures. 

\begin{figure}%[htbp]
\begin{center}
\includegraphics[width=0.75 \columnwidth]{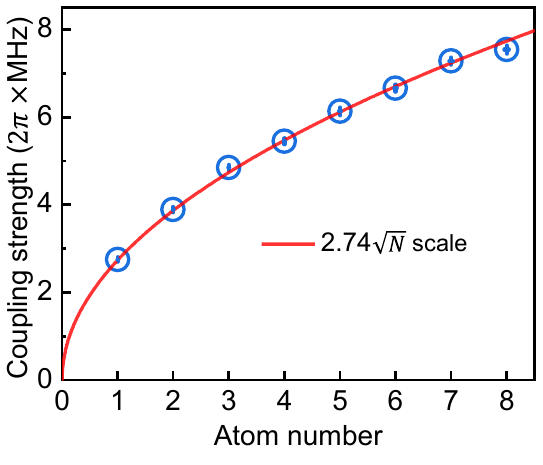}
\end{center}
\caption{\label{fig4} Dependence of the collective coupling strength on the atom number. The solid red line is the theoretical result for the collective enhancement relation ${\Omega _N} = g\sqrt N$ with the measured single atom coupling strength $g={\rm{2}}\pi  \times 2.74$ MHz.}
\end{figure}

The extracted vacuum Rabi splitting ${\Omega _N}$ versus atom number $N$ is displayed in Fig. \ref{fig4}. The single-atom coupling strength $g$ can also be deduced from ${\Omega _N}$ by $g = {{{\Omega _N}} \mathord{\left/
 {\vphantom {{{\Omega _N}} {\sqrt N }}} \right.
 \kern-\nulldelimiterspace} {\sqrt N }}.$
We obtained eight values of $g$ corresponding to atom numbers from 1 to 8. The variance in $g$ is within $\pm 2\%$ of the average value ${\rm{2}}\pi  \times 2.74$ MHz. The single-atom coupling strength can also be obtained by fitting the data with $g' = {{{\Omega _N}} \mathord{\left/
 {\vphantom {{{\Omega _N}} {\sqrt N }}} \right.
 \kern-\nulldelimiterspace} {\sqrt N }}
$, which gives $g' = {\rm{2}}\pi  \times 2.73(1)$ MHz, and it is almost the same as the averaged value.

Collective enhancement of light-matter interactions by using multiple single atoms is a basic principle in quantum physics. The dependence of the collective enhancement on the atom number has been proven through Rydberg excitation of atoms \cite{Dudin2012, Labuhn2016} and single qubits in superconducting circuits \cite{Wang2020}. Here, this fundamental relation can be tested by using deterministic atom numbers with discrimination at the real single-atom level in our cavity QED system. As displayed in Fig. \ref{fig4}, the theoretical collective enhancement relation $ {\Omega _N} = g\sqrt N$ is shown as the red line with the single-atom coupling strength $g={\rm{2}}\pi  \times 2.74$ MHz. We see that the scaling of experimental data with the atom number agrees very well with the theory, which validates the principle of collective enhancement. 

In summary, we have developed a new cavity QED system in which well-controlled 1D atom arrays are strongly coupled to a miniature F-P cavity. Vacuum Rabi splittings with a deterministic number of atoms are observed; thus, the principle of collective enhancement of light-matter interactions with multiple atoms is experimentally tested and validated with single atoms. The system provides a versatile platform to study light-matter interactions, quantum networks with nodes containing multiple atomic qubits \cite{Science.Kaufman.2021}, and many-body physics with interactions mediated by photons \cite{PhysRevLett.122.010405,Nature.Thompson.2020,Nature.Smith.2021, PhysRevLett.110.090402}.

\begin{acknowledgements}
This work was supported by the National Key Research and Development Program of China (Grans Nos. 2021YFA1402002 and 2017YFA0304502), the National Natural Science Foundation of China (Grant Nos. U21A6006, U21A20433, 11974223, 11974225, 12104277, and 12104278), and the Fund for Shanxi 1331 Project Key Subjects Construction.

\end{acknowledgements}


\bibliographystyle{stray}
\bibliography{bibl}

\pagebreak 
\clearpage

\widetext

\renewcommand{\thefigure}{S\arabic{figure}}
\setcounter{figure}{0}
\renewcommand{\thetable}{S\arabic{table}}
\setcounter{table}{0}
\renewcommand{\theequation}{S\arabic{equation}}
\setcounter{equation}{0}

\section{Supplementary Materials for ``Realization of strong coupling between deterministic single-atom arrays and a high-finesse miniature optical cavity''}

\subsection*{ Yanxin Liu, Zhihui Wang, Pengfei Yang, Qinxia Wang, Qing Fan, Shijun Guan, Gang Li, Pengfei Zhang, and Tiancai Zhang}

\subsection*{State Key Laboratory of Quantum Optics and Quantum Optics Devices,
and Institute of Opto-Electronics, Shanxi University, Taiyuan 030006, China\\
Collaborative Innovation Center of Extreme Optics, Shanxi University, Taiyuan 030006, China
}

\affiliation{State Key Laboratory of Quantum Optics and Quantum Optics Devices,
and Institute of Opto-Electronics, Shanxi University, Taiyuan 030006, China}
\affiliation{Collaborative Innovation Center of Extreme Optics, Shanxi University, Taiyuan 030006, China}

\section{Theory of the vacuum Rabi spectrum of a multiatom cavity QED}

For a dissipated multiatom cavity QED system with a near-resonance coherent laser at a frequency of ${\omega _p}$ as the pump light to the cavity, the total Hamiltonian is \cite{CQED2005}:
\begin{equation}
{H_{total}} = {H_{TC}} + {H_P}.
\end{equation}
The first term ${H_{TC}}$ represents the Tavis-Cummings Hamiltonian for a system of multiple atoms interacting with a single mode cavity
\begin{equation}
{H_{TC}} = \hbar {\omega _c}{a^ + }a + \hbar \sum\limits_{k = 1}^N {{\omega _k}\left( {{\sigma _k}^z + \frac{1}{2}} \right)}  + \hbar \sum\limits_{k = 1}^N {{g_k}\left( {{a^ + }{\sigma _k}^ -  + {\sigma _k}^ + a} \right)} ,
\end{equation}
and the second term is the pump term with the form 
\begin{equation}
{H_P} = \hbar \eta \left( {a{e^{i{\omega _p}t}} + {a^ + }{e^{ - i{\omega _p}t}}} \right).
\end{equation}
Here, ${\omega _c}$ is the resonance frequency of the cavity mode; ${\omega _k}$ is the transition frequency of the ${k^{th}}$ atom between the excited energy level $\left| e \right\rangle$ and the ground energy level $\left| g \right\rangle$; ${a^ + }$ and $a$  are the creation and annihilation operators of the photon; the atomic spin operators of ${\sigma _k}^ +  = \left| e \right\rangle \left\langle g \right|$ and ${\sigma _k}^ -  = \left| g \right\rangle \left\langle e \right|$ are ladder operators for the ${k^{th}}$ atom; $\sigma _k^z = \frac{1}{2}\left( {\left| e \right\rangle \left\langle e \right| - \left| g \right\rangle \left\langle g \right|} \right)$ is the Pauli $z$-operator for the ${k^{th}}$ atom; ${\sigma _k}^ + a + {\sigma _k}^ - {a^ + }$ represents the interaction of the cavity mode and the ${k^{th}}$ atom; $\eta$ denotes the strength of the probe beam; ${g_k}$ is the coupling strength of the ${k^{th}}$ atom to the cavity mode and $N$ is the effective number of atom coupled to the cavity mode.

By transforming to a reference frame associated with the probe beam and rewriting the Hamiltonians in terms of collective spin operators, we obtain
\begin{equation}
{H_{TC}} = \hbar {\Delta _{pc}}{a^ + }a + \hbar {\Delta _{pa}}{J^z} + \hbar \sqrt N g\left( {{J^ + }a + {a^ + }{J^ - }} \right)
\end{equation}
and
\begin{equation}
{H_P} = \hbar \eta \left( {a + {a^ + }} \right),
\end{equation}
where
\begin{equation}
{J^z} = \sum\limits_{k = 1}^N {\sigma _k^z}
\end{equation}
and
\begin{equation}
{J^ \pm } = \sum\limits_{k = 1}^N {\frac{{{g_k}}}{{\sqrt {\sum\limits_{k = 1}^N {{{\left| {{g_k}} \right|}^2}} } }}} \sigma _k^ \pm  = \frac{1}{{\sqrt N }}\sum\limits_{k = 1}^N {\sigma _k^ \pm } 
\end{equation}
are the atomic collective spin operators. The frequency detuning between the probe light and the atoms is ${\Delta _{pa}} = {\omega _p} - {\omega _a}$. The frequency detuning between the probe light and cavity is ${\Delta _{pc}} = {\omega _p} - {\omega _c}$. The ${\omega _k}$ and ${g _k}$ for different atoms are assumed to be the same, respectively, as ${\omega _k}=\omega$ and ${g _k}=g$.

The dynamics of the system can be theoretically obtained by solving the Lindblad master equation
\begin{equation}
\dot \rho  =  - \frac{i}{\hbar }\left[ {{H_{TC}} + {H_P},\rho } \right] + {\cal L}\left[ \rho  \right].
\end{equation}
${\cal L}\left[ \rho  \right]$ is the Liouville superoperator, which describes the coupling of the cavity mode and the atoms to the environment, and it can be expressed as
\begin{equation}
{\cal L}\left[ \rho  \right] = \kappa \left( {2a\rho {a^ + } - {a^ + }a\rho  - \rho {a^ + }a} \right) + \gamma \sum\limits_{k = 1}^N {\left( {2{\sigma _k}^ - \rho {\sigma _k}^ +  - {\sigma _k}^ + {\sigma _k}^ - \rho  - \rho {\sigma _k}^ + {\sigma _k}^ - } \right).}
\end{equation}
Thus, the time evolution for the expectation value of any operators $\hat o$ can be calculated by \cite{Dombi2013,Howard1991}
\begin{equation}
\left \langle  \dot{\hat{o} }   \right \rangle   = Tr\left[ {\hat o\dot \rho } \right]
\end{equation}

The time evolutions for $\langle a \rangle$, $ \langle {J^ - } \rangle $ and $\langle {J^z} \rangle$ can be obtained as
\begin{equation}
\left\langle {\dot a} \right\rangle  = i\left( {{\Delta _{pc}} + i\kappa } \right)\left\langle a \right\rangle  - ig\sqrt N \left\langle {{J^ - }} \right\rangle  - i\eta ,
\end{equation}
\begin{equation}
\left\langle {{{\dot J}^ - }} \right\rangle  = i\left( {{\Delta _{pa}} + i\gamma } \right)\left\langle {{J^ - }} \right\rangle  + \frac{{i2g}}{{\sqrt N }}\left\langle {{J^z}a} \right\rangle
\end{equation}
and
\begin{equation}
\left\langle {{{\dot J}^z}} \right\rangle  =  - 2\gamma \left( {\frac{N}{2} + \left\langle {{J^z}} \right\rangle } \right) + 2i\sqrt N g\left\langle {{a^ + }{J^ - } - {J^ + }a} \right\rangle ,
\end{equation}
respectively.

When the probe laser is very weak, it can be assumed that only one atom can be excited and all the others remain in the ground state. Thus, the system evolves within the Hilbert space spanned by states$\left\{ \left| G \right\rangle \otimes \left| 0 \right\rangle, \\ \left| G \right\rangle \otimes \left| 1 \right\rangle, \\ \left| E \right\rangle \otimes \left| 0 \right\rangle \right\}$, with $\left| G \right\rangle  = \left| {gg \cdots gg} \right\rangle$ and  $\left| E \right\rangle  = \frac{1}{{\sqrt N }}\left( {\left| {eg \cdots gg} \right\rangle + \left| {ge \cdots gg} \right\rangle + \cdots + \left| {gg \cdots ge} \right\rangle } \right)$ being the collective atom states. We then have $\left\langle {{J^z}} \right\rangle  =  - \frac{N}{2}$ and $\left\langle {{J^z}a} \right\rangle  =  - \frac{N}{2}\left\langle a \right\rangle$. Therefore,  Eqs (S11) and (S12) can be simplified to
\begin{equation}
\left\langle {\dot a} \right\rangle  = i\left( {{\Delta _{pc}} + i\kappa } \right)\left\langle a \right\rangle  - i{\Omega _{eff}}\left\langle {{J^ - }} \right\rangle  - i\eta
\end{equation}
and
\begin{equation}
\left\langle {{{\dot J}^ - }} \right\rangle  = i\left( {{\Delta _{pa}} + i\gamma } \right)\left\langle {{J^ - }} \right\rangle  - i{\Omega _{eff}}\left\langle a \right\rangle ,
\end{equation}
where ${\Omega _{eff}} = \sqrt N g$. The steady state of the system can be obtained by setting $\left\langle {\dot a} \right\rangle  = 0$, $\left\langle {{{\dot J}^ - }} \right\rangle  = 0$. The transmission spectrum of the coupled system can be finally obtained as  
\begin{equation}
T = \frac{{{\kappa ^2}\left( {{\gamma ^2} + \Delta _{pa}^2} \right)}}{{{{\left( {{\Omega _{eff}}^2 - \Delta _{pa}^2 + {\Delta _{ca}}{\Delta _{pa}} + \gamma \kappa } \right)}^{^2}} + {{\left( {\kappa {\Delta _{pa}} + \gamma {\Delta _{pa}} - \gamma {\Delta _{ca}}} \right)}^{^2}}}},
\end{equation}
in which ${\Delta _{ca}} = {\omega _c} - {\omega _a}$ is the frequency detuning between the cavity and atom.

\section{The double MOT vacuum system}
\begin{figure}
\centering{}\includegraphics[width=0.8 \columnwidth]{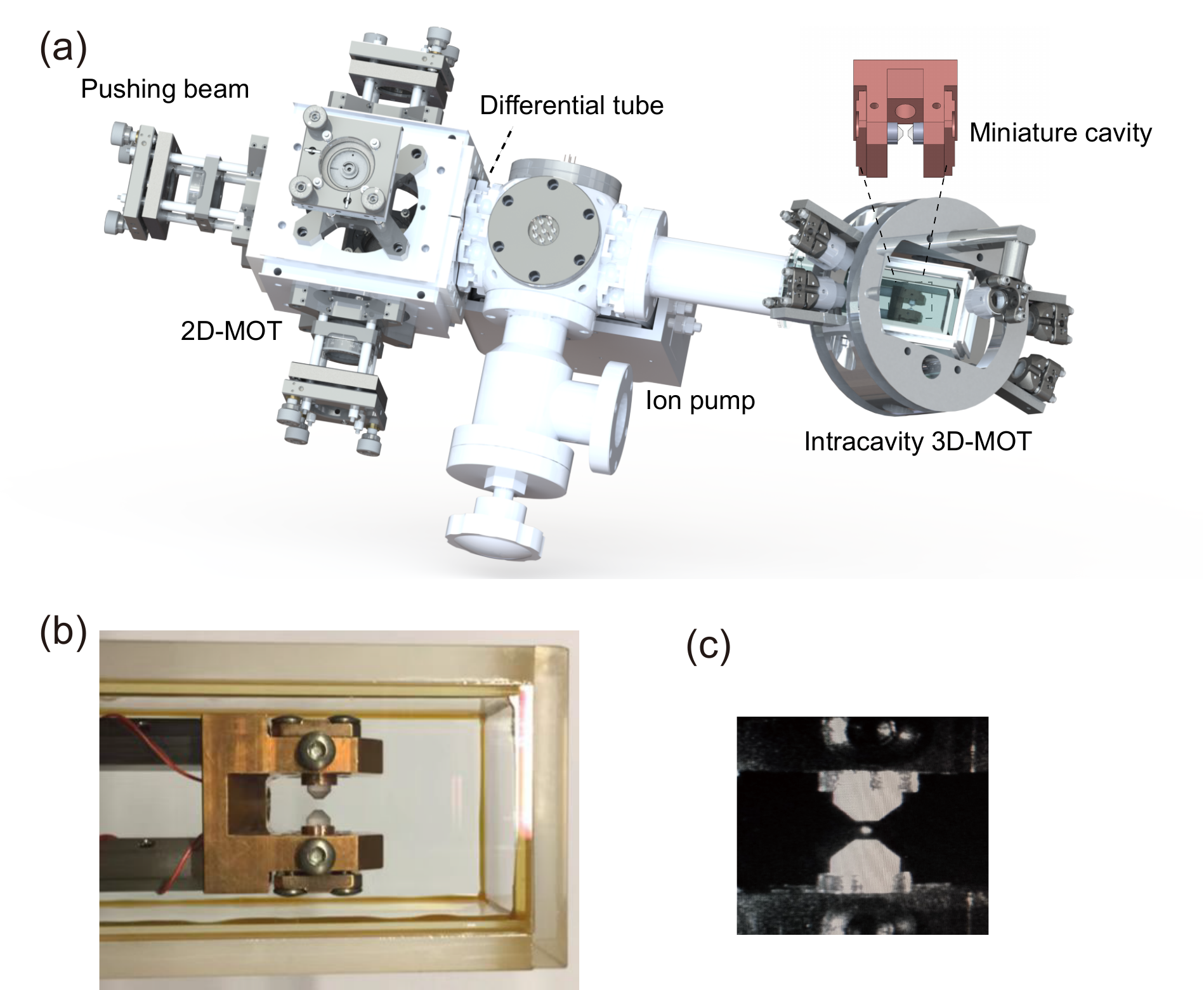} \caption{\label{figS1} (a) Schematic of the vacuum chamber associated with the optics for MOT system. (b) Picture of the miniature cavity and mount residing in the vacuum glass cell. (c) Fluorescence of the 3D MOT in the miniature cavity.
 }
\end{figure}

The drawing of the whole vacuum system is shown in Fig. \ref{figS1}(a), which includes a two-stage Magnetically Optical Trap (MOT) system. The first stage is a two-dimensional (2D) MOT, and the second stage is a three-dimensional MOT. The two parts are separated by a differential tube with an inner diameter of 1 mm and length of 10 mm. The light beams of the 2D-MOT are shaped as ellipses with a size of approximately 6 mm $\times$ 12 mm. The optical power of each cooling beam and the repumping beam is 20 mW and 2 mW, respectively. The cesium atoms emitted from the dispenser are precooled by the 2D MOT and pushed to the 3D MOT with the aid of a pushing beam. The 3D-MOT resides in the center of the miniature Fabry--Perot (F-P) cavity with a length of 1.27 mm and captures the atoms from the 2D-MOT. To accommodate the 3D MOT within the cavity, the MOT beams are also converted to an ellipse with a size of approximately 0.8 mm $\times$ 1.6 mm. The angle between the transverse beams for the MOT is also squeezed to ${20^ \circ }$. The powers of each cooling and repumping beam are 400 $\mu$W and 60 $\mu$W, respectively. Eventually, we obtain approximately ${10^5}$ atoms in the 3D MOT. The MOT coils and the optical holders are fixed together on a single mount; thus, the position of the 3D MOT can be moved freely and finely by a 3D positioner.

Figure \ref{figS1}(b) shows a picture of the F-P cavity, and the mount resides in a high-vacuum glass cell with inner dimensions of 20 mm $\times$ 25 mm on the cross section. Figure \ref{figS1}(c) displays the fluorescence of the atomic ensemble captured in the 3D MOT.

\section{Scheme of cavity locking}
\begin{figure}
\centering{}\includegraphics[width=0.6 \columnwidth]{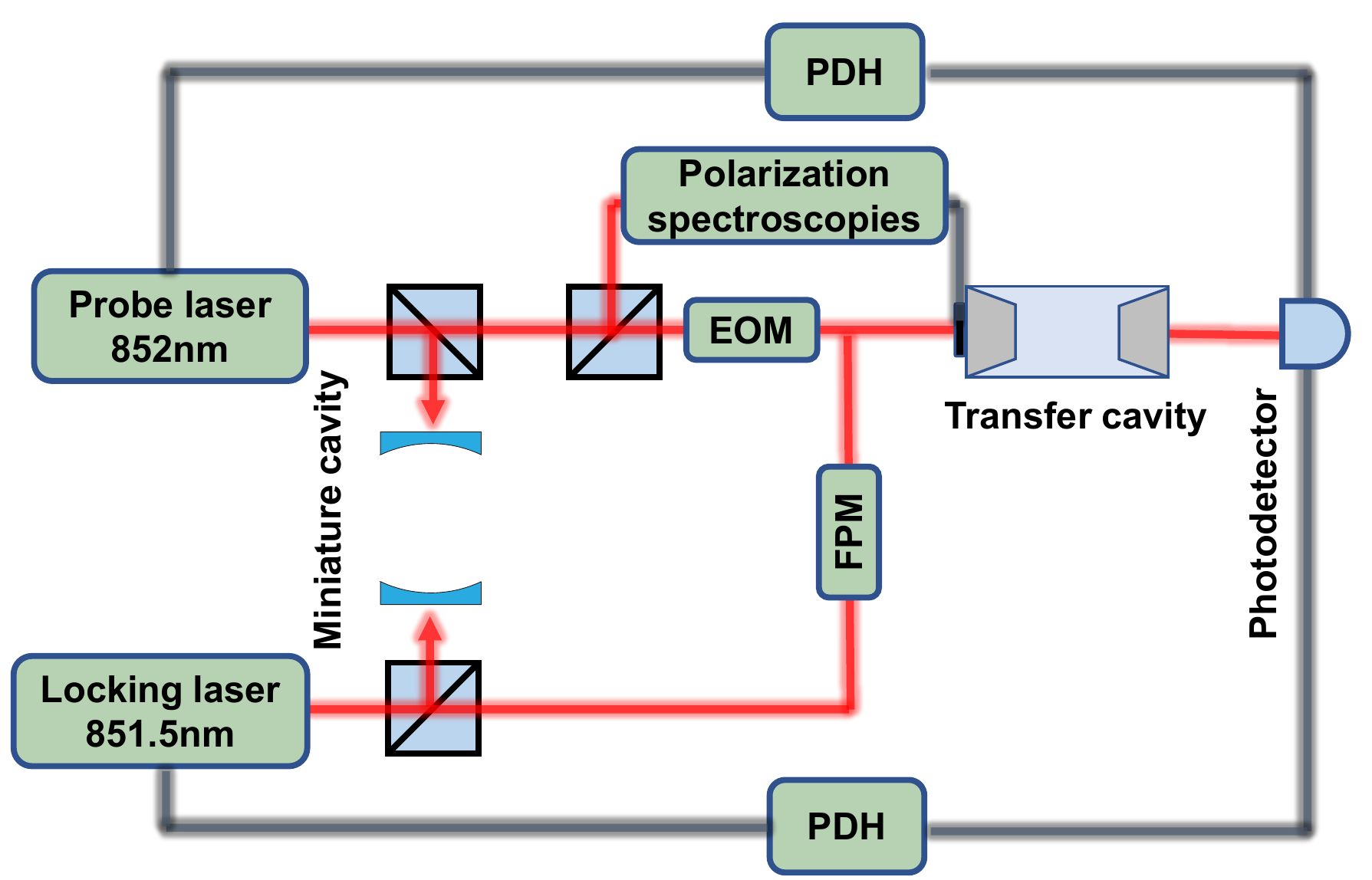} \caption{\label{figS2} The frequency locking scheme for the miniature cavity.
Both the probe and locking lasers are homemade external cavity diode lasers (ECDLs). FPM: fiber phase modulator (Photline NIR-MPX800), EOM: electro-optic modulator, PDH: Pound-Drever-Hall.   }

\end{figure}

The miniature F-P cavity is locked by an 851.5-nm locking laser that is resonant to the F-P cavity. The frequency is blue detuned to the 852-nm mode by 354.3 GHz, which is 3 free spectral ranges of the cavity away from 852-nm. The frequency of the locking laser is stabilized to a transfer cavity (TC) which is prelocked to cesium transition lines via the 852-nm probe laser. The scheme of the locking system is shown in Fig. \ref{figS2}. 

The probe laser, which is from an external cavity diode laser, is first locked to the TC with a linewidth of 1 MHz by the Pound-Drever-Hall (PDH) method \cite{Black2001} to suppress the linewidth. The long-time frequency drift is then suppressed by locking the whole system, which includes both the probe laser and TC, to the cesium transition line. The cesium transition line is resolved with polarization spectroscopy. After these two steps, a narrow-linewidth and stable probe laser and a frequency-stabilized TC cavity are obtained. The frequency of the 851.5-nm locking laser is then stabilized to the TC by another PDH locking loop. Finally, the locking laser is fed to the miniature F-P cavity via a fiber phase modulator, and the cavity is locked to one of the sidebands. By controlling the frequency of the applied rf signal to the fiber phase modulator, the resonance frequency of the miniature cavity can be tuned freely. The standing-wave field of the locking laser inside the miniature cavity also forms a blue detuned lattice to confine the cesium atoms in the Z direction.

\section{Analysis of the consistency of the optical tweezers}
\begin{figure}
\centering{}\includegraphics[width=0.8 \columnwidth]{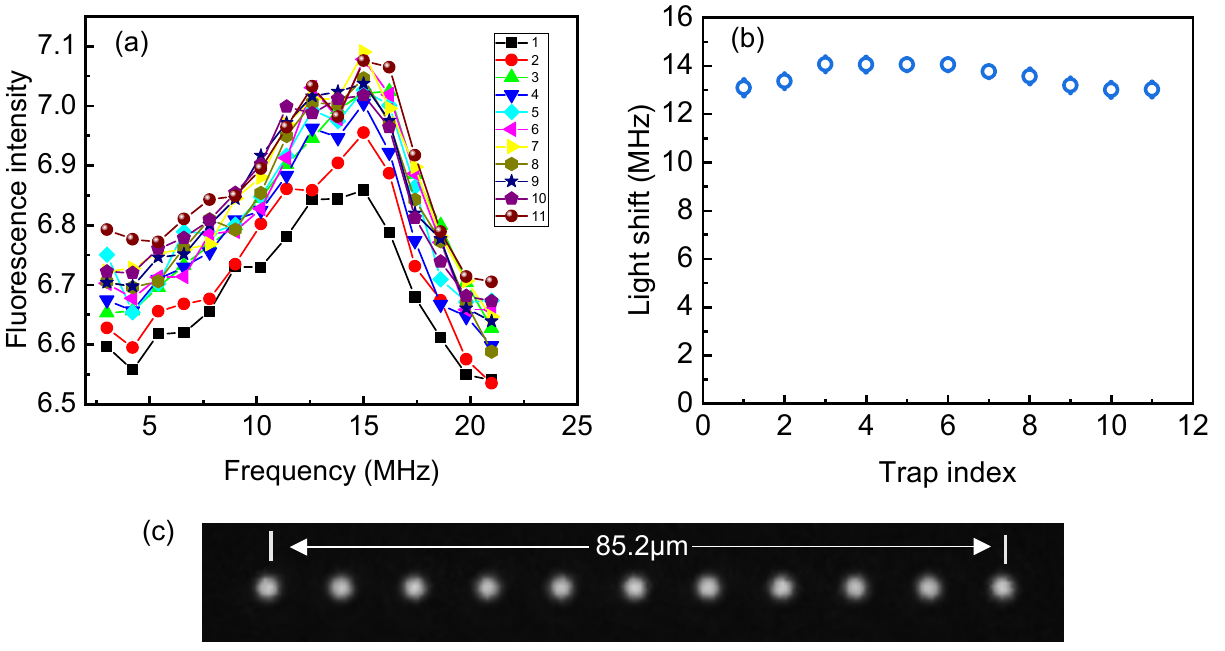} \caption{\label{figS3}Characterization of the dipole trap array. (a) is the fluorescence spectra of the single atom trapped in the dipole tweezer array, and (b) is the extracted light shift for every tweezer in the array, which is obtained by fitting the spectra in (a). The error bar in (b) is the fitting error. (c) is the image of the final dipole trap array magnified by 8.27 times. The size of each pixel is 3.45 $\mu$m. 
}

\end{figure}

As we presented in the main text, the optical tweezer array is generated by focusing 11 optical beams at 1064 nm from diffraction of an acousto-optic deflector (AOD) with a homemade objective. The consistency of the trap mainly depends on the diffraction performance of the AOD. However, the intermodulation between different RF tones and nonlinearity of the diffraction would also cause the inhomogeneity. The inhomogeneity can be suppressed by adjusting the phases and amplitude of the different RF tones \cite{Manuel2016}. To characterize the consistency of the optical tweezers, we have measured the fluorescence spectrum of single atoms trapped in every optical tweezer by scanning the frequency of the imaging beam. The intensity of fluorescence is recorded by the EMCCD camera in this process. The light shift induced by the 1064-nm trap beam can be extracted from the spectra. The results of the fluorescence spectra and the extracted light shift for every single atom in each tweezer are shown in Fig. S3(a) and (b). The variance of the light shifts for all 11 tweezers is within 3.2\%, which also reflects the intensity inhomogeneity of the tweezers. The distance between adjacent traps is determined to be 8.52 $\mu$m by directly imaging the trap array pattern on a CCD camera, which is shown in Fig. \ref{figS3}(c). The corresponding RF frequency interval is 1.16 MHz.

\section{The experimental process}
The time sequence of the experiment is shown in Fig. \ref{figS4}(a). The main steps include 1) MOT loading in the first 780 ms. Cold atoms in a 2D MOT are accumulated from the background vapor and then pushed through a differential tube with a diameter of 1 mm to the center of the miniature cavity. A small cold atomic cloud is captured in the center of the cavity by a 3D MOT. At the end of the MOT loading period, the tweezer array is switched on and overlaps with the atomic cloud for 100 ms to load the atoms into the tweezers. During the loading phase, the barrier of the intracavity lattice is set as approximately 11 $\mu$K. 2) Two-stage PGC in 12 ms. The quadrupole magnetic field for the MOT is turned off, and then two polarization gradient cooling (PGC) phases, where the frequency detunings of the cooling beam are $-30$ MHz and $-60$ MHz, are applied sequentially to cool the atoms. The final temperature is approximately 15 $\mu$k. 3) Light-assisted collision in 20 ms. An extra 20-ms light assisted collision phase induced by the imaging beam is deliberately applied to ensure that single atom is loaded in every single tweezer. 4) Second PGC in 6 ms. Another PGC phase with $-30$-MHz frequency detuning of the cooling beam  is applied to cool the atom again. 5) State preparation in 2 ms. After the single cesium atoms are successfully loaded into the tweezers, the atom is then initialized into the Zeeman sublevel $ 6{S_{1/2}}|F = 4,{m_F} = 4\rangle $ by scanning the frequency of the cavity probe laser with ${\sigma ^ + }$ polarization. 6) Measurement of the vacuum Rabi splitting (VRS) in 1 ms. Before the VRS measurement, the trap depth of the 1064-nm tweezers are decreased to $k_B \times$0.1 mK in 5 ms, and meanwhile, the trap barrier of the lattices is increased to $k_B \times$0.29 mK. Then, the VRS is obtained by scanning the probe frequency in 1 ms. Afterwards, the trap depths of both the optical tweezers and the latiices ramping back in 2 ms. 6) EMCCD imaging in 40 ms. The atom number in the trap array is then determined by the fluorescence image of EMCCD.

\subsection{Details of the atom state preparation}

The single atoms trapped in the tweezer array are initialized into the Zeeman sublevel $ 6{S_{1/2}}|F = 4,{m_F} = 4\rangle $ by a probe light with $\sigma^+$-polarization. However, due to the intrinsic birefringence of the high-finesse cavity, the polarization of the cavity field would be depolarized and depresses the efficiency of the atom state preparation. To increase the efficiency, the intracavity field is deliberately optimized to a highly polarized $\sigma^+$ by examining the transmitting field. The frequency of the probe light is then scanned to change $\Delta_{pc}$ from $-20$ to 20 MHz in 2 ms to realize the state initialization.

\begin{figure}
\centering{}\includegraphics[width=0.8 \columnwidth]{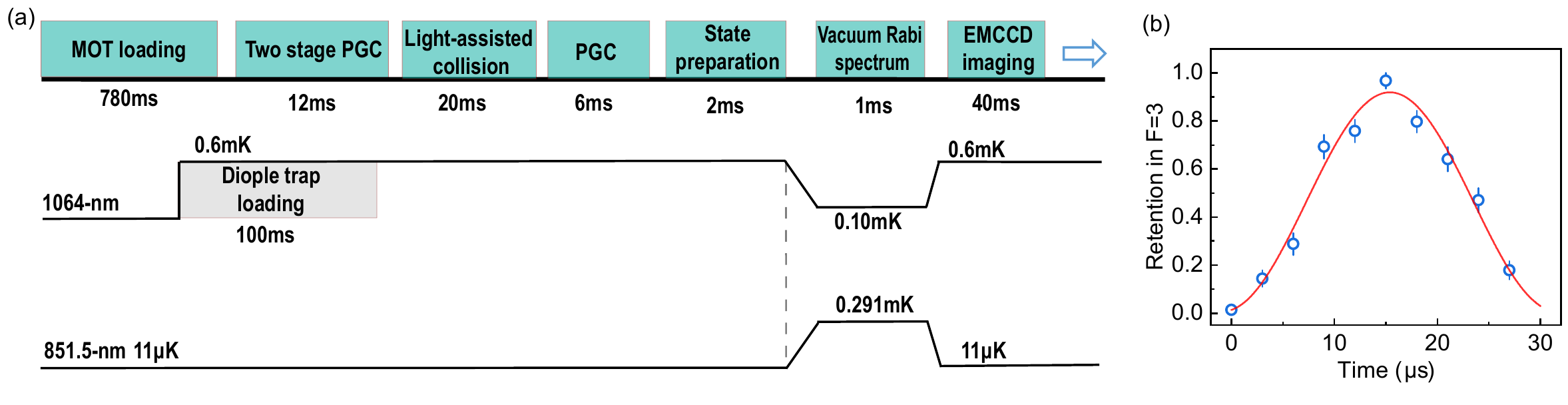} \caption{\label{figS4} (a) Time sequence of the experimental process. (b) Rabi oscillation between states $ 6{S_{1/2}}|F = 4,{m_F} = 4\rangle $ and $ 6{S_{1/2}}|F = 3,{m_F} = 3\rangle $. The atom is initially prepared in state $ |F = 4,{m_F} = 4\rangle $, a microwave pulse transfers the atom to $ |F = 3,{m_F} = 3\rangle $, the atom in $ |F = 4\rangle $ is blown away, then the retention in$ |F = 3,{m_F} = 3\rangle $ is measured, and a sinusoidal oscillation is finally observed. We deduce that the efficiency of state initialization is greater than 91.2\% from the fitted amplitude.  
}
\end{figure}

The efficiency of the state preparation is evaluated by the amplitude of Rabi flopping between $ 6{S_{1/2}}|F = 4,{m_F} = 4\rangle $ and $ 6{S_{1/2}}|F = 3,{m_F} = 3 \rangle $, which is driven by microwaves. The typical Rabi flopping signal is given in  Fig. \ref{figS4}(b). The maximum efficiency with which the atom is transferred to $ 6{S_{1/2}}|F = 3,{m_F} = 3 \rangle $ is 96.7\%, and data fitting with a sine function gives a maximum transfer efficiency of 91.2\%. This indicates that the efficiency of the state preparation is above 91.2\%.

\subsection{Imaging the single atoms in the trap array}
The number of atoms coupling with the cavity is determined by fluorescence imaging of the single atoms in the tweezer array. A weak retroreflected imaging beam with a diameter of 400 $\mu m$ and orientation of $45 ^{\circ}$ in the X-Y plane is used to excite the atom. The imaging beam is linearly polarized and passes through the center of the cavity. Then, it is reflected back through a $ \lambda/4$ wave plate. Thus, a lin$ \bot$ lin configuration is formed. This beam is red detuned by 24 MHz from the atomic transition line $ 6{S_{1/2}}|F = 4\rangle \leftrightarrow 6{P_{3/2}}|F = 5\rangle $. Therefore, it can also be used to cool atoms during imaging process. During this process, the miniature cavity is dragged to 10 MHz red detuned from the atomic transition $ 6{S_{1/2}}|F = 4\rangle \leftrightarrow 6{P_{3/2}}|F = 5\rangle $ to enhance the atomic radiation rate in the transverse direction of the cavity. 

\section{Error estimation of measured coupling strengths and atom numbers}
\subsection{Estimation of the variance of the atom-cavity coupling strength due to thermal motion of the atom}
We will estimate the variance in the atom-cavity coupling strength due to the thermal motion of the atom by using the following experimental conditions: the temperature of the trapped atoms is $T = $ 15 $\mu$K and the depth of the blue-detuned intracavity lattice is ${U_0} = k_B \times 0.29$ mK. The atom motion in the lattice along the Z axis can be approximated by a quantum harmonic oscillator with oscillating frequency $\omega_z=\frac{2 \sqrt{2} \pi}{\lambda_\text{lattice}} \sqrt{\frac{U_0}{m}}$ with $\lambda_\text{lattice}$ and $m$ being the wavelength of the lattice beam and mass of a cesium atom, respectively. The wavefunction on a state with quantum number $n$ is:
\begin{equation}
|\Psi\rangle_{n}=\sqrt{ \frac{1}{n! 2^n a \sqrt{\pi}} } \times \text{H}_n [(z-z_0)/a] e^{-\frac{(z-z_0)^2}{2 a^2}},
\end{equation}
where $\text{H}_n (z)$ are the Hermite polynomials, $z_0$ is the coordinate where the atom is trapped and $a=\sqrt{\frac{\hbar}{m \omega_z}}$. The average phonon number for a thermal atom with temperature $T$ can be approximately obtained by $\langle n \rangle=\frac{k_B T}{\hbar \omega_z}=1.3$. Thus, the probability of the atom on state with phonon number $n$ is $P_n= \langle n \rangle^n/(1 + \langle n \rangle)^{(n + 1)}$, which is also a Bose distribution. 

We assume that the atoms in the trap array are placed in the center of the cavity mode along the X and Y axes. The coupling strength along the Z direction satisfies the relation $g=g_0 |\cos(2kz)|$ with $ k= {2\pi}/\lambda$ the wave vector, $\lambda$ 852.3 nm, and maximum coupling strength $g_0= 2\pi \times 3.16$ MHz, which is obtained theoretically by the geometry of the cavity. The coupling strength due to the thermal distribution of the atom can be calculated by 
\begin{equation}
g_\text{a}=\sum^{n}\int g P_n  \langle \Psi |\Psi\rangle_{n} \text{d}z. 
\end{equation}
Because the atoms are mainly populated in the low phonon state, we can truncate the sum of the phonon number at 10. Due to the small wave package of the wavefunction the integrating range can also be truncated at $\lambda_\text{lattice}/4$, and the integrating range is then $(z_0-\lambda_\text{lattice}/4,z_0+\lambda_\text{lattice}/4)$. At the position where the node of the lattice perfectly overlaps with the antinode of the 852.3-nm cavity mode, the average coupling strength due to the thermal distribution of atoms is $ 2\pi \times 3.1$ MHz. At the position where the node of the lattice perfectly overlaps with the node of the 852.3-nm cavity mode, the average coupling strength due to the thermal distribution of atoms is $ 2\pi \times 0.45 $ MHz, which is 14.2\% of $g_0$.

\subsection{The error estimation of the atom number}
As presented in Section V, the atom number is determined after the measurement of the VRS. The time from the beginning of the VRS measurement to atom number measurement is 3 ms. By taking into account the single-atom lifetime of 4.8 s, we obtain that the loss probability of one atom is only 0.06\%. In an array of N atoms, the survival probability of all the atoms will be ${0.9994}^N$. In our experiment, the maximum atom number is 8, and the survival probability is still 99.5\%. The reading error of a single atom is $8.9\times{10}^{-7}$, which is estimated from the data shown in Fig. 2(a). Therefore, the reading fidelity for 1 to 8 atoms is almost equal to 1 and the errors can be omitted. As an overall estimation, the errors of the atom number for 1 to 8 are mainly caused by the atom's lifetime. The errors are 0.0006, 0.0024, 0.0054, 0.0096, 0.015, 0.0216, 0.0293, and 0.0383 for 1 to 8 atoms, respectively, and they are shown as error bars of atom numbers in Fig. 4.

\end{document}